%% file: main.tex
\documentclass[]{article}
\usepackage[margin=1in]{geometry}
\usepackage[utf8]{inputenc}
\usepackage{paralist}
\usepackage{textcomp}

\title{Revisiting hBFT: Speculative Byzantine Fault Tolerance with Minimum Cost}
\author{ \parbox{3.5 in}{\centering Nibesh Shrestha, Mohan Kumar\\
        Department of Computer Science\\
        Rochester Institute of Technology, NY, USA
        {\tt\small \{nxs4564, mjkvcs\}@rit.edu}
         \\with:\\
        SiSi Duan\\
        University of Maryland, Baltimore\\
        {\tt\small sduan@umbc.edu}
        }
}
\date{}
\begin{document}

\maketitle
\begin{abstract}
    FaB Paxos\cite{martin2006fast} sets a lower bound of $5f+1$ replicas for any two-step consensus protocols tolerating $f$ byzantine failures. Yet, hBFT\cite{duan2015hbft} promises a two-step consensus protocol with only $3f+1$ replicas. As a result, it violates safety property of a consensus protocol. In this note, we review the lower bound set by FaB Paxos and present a simple execution scenario that produces a safety violation in hBFT. To demonstrate the scenario, we require a relatively simple setup with only 4 replicas and one view-change.
\end{abstract}

\input{sections/introduction}
\input{sections/preliminaries}
\input{sections/skeletal_overview_fab}
\input{sections/skeletal_overview}
\input{sections/violation}

\bibliographystyle{abbrv}
\bibliography{main}
\end{document}

%% file: sections/introduction.tex
\section{Introduction}
\label{sec:Introduction}
A byzantine fault tolerant consensus protocol involves a set of replicas $N$ to reach an agreement on a common value among correct replicas in the presence of $f$ malicious replicas. Fast byzantine consensus protocols such as FaB Paxos\cite{martin2006fast} requires only two communication steps to reach consensus in the common case. A common case execution consists of a (possibly honest) proposer replica proposing to all replicas among which at most $f$ replicas could be byzantine faulty. In this setup, FaB Paxos requires at least $5f+1$ replicas and sets a lower bound of $5f+1$ replicas for any two-step consensus protocols tolerating $f$ byzantine failures. In the same common case setup, hBFT\cite{duan2015hbft} promises a two-step consensus protocol with an optimal (i.e. $3f+1$) number of replicas. However, this violates the lower bound set by FaB Paxos and as a result, the protocol fails to guarantee safety property of a consensus protocol.

In this note, we review the lower bound of $5f+1$ replicas required for any two-step byzantine fault tolerant consensus protocol and present a simple scenario in which a single faulty primary can break safety of hBFT protocol. In \cite{abraham2017revisiting}, Abraham et al. show a similar safety violation in Zyzzyva \cite{kotla2007zyzzyva}. Zyzzyva also requires $3f+1$ replicas to reach consensus in two communication steps. However, Zyzzyva's common case execution involves an optimistic execution where no failure occurs and all $3f+1$ replicas respond identically. This does not constitute a violation of the lower bound for a common case execution with at most $f$ byzantine failures. The safety violation in Zyzzyva is a result of incorrectly selecting a possibly committed value during view-change; a process by which a new leader replica is selected. In \cite{abraham2018revisiting}, Abraham et al. propose a correct solution to fix the issue.

%% file: sections/preliminaries.tex
\section{Preliminaries}
\label{sec:Preliminaries}
In this note, we consider byzantine fault tolerant consensus protocols involving a group of replicas among which at most $f$ replicas can suffer byzantine failures. The replicas exchange messages to agree on a common value. The communication channel is authenticated, reliable and asynchronous; messages sent between replicas are never lost, but may take a long time before they finally arrive. In the consensus protocol, one replica is chosen as the \textit{primary} and other replicas are \textit{backups}. The \textit{primary} is responsible for proposing values to other replicas. A value is said to be committed when a quorum (usually $N-f$) of replicas have accepted a common value.

The consensus protocol should satisfy following properties:
\begin{itemize}[]
    \item \textbf{Agreement.} All correct replicas commit on a common value. This property is also called \textbf{safety}. 
    \item \textbf{Validity.} A value committed by a correct replica must be proposed by a \textit{primary}.
    \item \textbf{Termination.} A value proposed by a \textit{primary} must eventually be committed provided the communication channel is eventually partially synchronous.
\end{itemize}

FaB Paxos is a generic consensus protocol designed only for reaching agreement. It separates the roles of replicas into \textit{proposers}, \textit{acceptors} and \textit{learners}. The \textit{proposer} is analogous to the \textit{primary} and the \textit{acceptors} and \textit{learners} are analogous to the \textit{backups}. In contrast, hBFT is a full state machine replication (SMR) protocol--involving both agreement and execution of proposed values and classify replicas into only \textit{primary} and \textit{backups}. For brevity, we adopt a common convention of classifying replicas into \textit{primary} and \textit{backups} and concern only with agreement phase of the protocol.

A view represents the system state with a distinct \textit{primary}. Views are numbered by view numbers. In a view $v$, the \textit{primary} proposes a value to other replicas via PREPARE messages. A replica responds to the PREPARE message by sending COMMIT messages to all other replicas. We use a common convention of PREPARE and COMMIT messages in exploring both protocols. When a replica fails to collect a quorum of COMMIT messages within a certain timeout interval, it triggers a view-change sub-protocol. A View-change sub-protocol is a common technique employed to elect a new \textit{primary} and ensure progress. In view-change sub-protocol, replicas send values sent by the primary of view $v$. The view-change sub-protocol must ensure that a value committed at a correct replica in view $v$ stays committed even in the new view $v+1$. 

%% file: sections/skeletal_overview_fab.tex
\section{Skeletal Overview of FaB Paxos}
\label{sec:skeletal_overview_fab}
FaB Paxos is an easy two-step consensus protocol. It requires a total of $N=5f+1$ replicas to reach agreement in the common case. In a view $v$, the \textit{primary} proposes a value $m$ to other replicas by sending PREPARE messages. Replicas accept the PREPARE message if they haven't already accepted other values for view $v$. If they accept the PREPARE message for value $m$, they reply to all other replicas by sending COMMIT messages. A value $m$ is effectively committed at view $v$ when $3f+1$ correct replicas have accepted the value $m$ at view $v$. A correct replica considers a value $m$ committed when it receives $N-f$ (i.e. $4f+1$) COMMIT messages for value $m$ at the same view.

A replica may fail to collect $N-f$ COMMIT messages for a value $m$ within certain timeout interval either because of a faulty \textit{primary} has sent PREPARE messages for different values to different replicas or because of the inherent asynchrony in the communication channel. In either case, the replica initiates a view-change in which it sends \textit{signed} copy of its most recent accepted value to the new \textit{primary} of view $v+1$. The new \textit{primary} waits for only $4f+1$ signed responses from the \textit{backups} as $f$ byzantine faulty replicas may not respond. With $4f+1$ signed responses, the new leader constructs a \textit{progress certificate} which serves as a proof to identify a possibly committed value. 

A \textit{progress certificate} vouches for a value $m$ if there is no other value $m'$ that appears at least $2f+1$ times in the \textit{progress certificate}. The new \textit{primary} proposes the value $m$ vouched by the \textit{progress certificate} in the new view $v+1$ along with the \textit{progress certificate}. Replicas change their accepted value to the value $m$ proposed by the new \textit{primary} if the \textit{progress certificate} vouches for the value $m$. With eventual synchrony, a (possibly correct) \textit{primary} will propose same value to all replicas and all correct replicas will accept the same value common value. The protocol completes when all correct replicas have accepted the same value and send COMMIT messages for the value.

\subsection*{Informal Sketch of Lower Bound}
Assume $A$ be the minimum number of replicas required. An asynchronous consensus protocol tolerating $f$ byzantine faults may wait for only $A-f$ replies in any step as $f$ byzantine faulty replicas may not reply. However, $f$ replicas, whose replies weren't received, may be correct and only the communication channel is slow. As a result, the replies from $f$ byzantine faulty replicas may still be included in $A-f$ replies that a correct replica collects.

A correct replica considers a value $m$ committed when it receives $A-f$ identical COMMIT messages for the value $m$. Out of these $A-f$ replicas that accepted value $m$, only $A-2f$ replicas may be correct and at most $f$ replicas could be byzantine faulty and may change their decision later. $f$ other replicas whose COMMIT messages aren't included in $A-f$ COMMIT responses could be correct and might have accepted a different value $m'$. At this stage, few replicas may  not receive required $A-f$ COMMIT responses in a timely manner triggering a view-change.

A consensus protocol must ensure that a value once committed stays committed at all future times. During view-change, the new \textit{primary} collects $A-f$ \textit{signed} copies of accepted values from the replicas. These $A-f$ \textit{signed} responses could contain responses from $f$ correct replicas that accepted value $m'$ and $f$ byzantine faulty replicas that could equivocate and change their accepted value to $m'$. In total, there could be $2f$ votes for value $m'$. To ensure a committed value stays committed, the number of votes for value $m$ must be more than $2f$, the minimum being $2f+1$. Hence, $A-f= 2f + 2f + 1$, (i.e $A = 5f+1$)

%% file: sections/skeletal_overview.tex
\section{Skeletal Overview of hBFT}
\label{sec:skeletal_overview_hBFT}
hBFT is a recent addition to byzantine fault tolerant SMR protocol that speculatively executes operation $op$ specified in a proposed value $m$ before the value $m$ is committed. It consists of four sub-protocols--\begin{inparaenum}[(i)]
\item agreement
\item checkpoint
\item view-change
\item client suspicion.
\end{inparaenum} We review only the agreement and view-change sub-protocols to show the safety violation.

A SMR protocol assigns a distinct sequence number $n$ to a value $m$ such that the pair $(n, m)$ is consistent among all correct replicas. For a given sequence number $n$ and view $v$, the \textit{primary} proposes a value $m$ by sending PREPARE messages for it. Replicas accept the value $m$ if they haven't accepted other values for $n$. Each replica sends COMMIT messages for value $m$ to all other replicas. A correct replica considers a value $m$ committed at $n$ when it receives $2f+1$ identical COMMIT responses from other replicas (including itself). This set of $2f+1$ COMMIT messages for value $m$ forms a \textit{commit certificate}. The \textit{commit certificate} for value $m$ at some sequence number $n$ serves as a proof that a value has been committed at $n$. 

A replica initiates a view-change when it fails to receive $2f+1$ COMMIT messages within certain time duration or when it receives $f+1$ COMMIT message for a different value $m'$ than the one it received in PREPARE message. To initiate a view-change, it sends VIEW-CHANGE messages to all replicas. The view-change sub-protocol of hBFT differs from that of FaB Paxos-- in hBFT, replicas send not only their recently accepted value, but also a \textit{commit certificate} (if any) in the VIEW-CHANGE message. A correct replica can initiate view-change when it receives $f+1$ VIEW-CHANGE messages from other replicas. The new \textit{primary} collects $2f+1$ VIEW-CHANGE messages before initiating a new view $v+1$. This set of $2f+1$ VIEW-CHANGE messages serves as a \textit{progress certificate} in hBFT. 

To ensure that a value committed in an old view stays committed even in the new view, the new \textit{primary} must choose possibly committed values based on the \textit{progress certificate} and (possibly) re-propose them. During view-change, the new \textit{primary} selects a value $m$ if there is at-least one \textit{commit certificate} for $m$ or if there are at least $f+1$ replicas who have accepted value $m$; or else NULL value is selected. The new \textit{primary} proposes the selected value $m$ in the new view $v+1$ along with the \textit{progress certificate} for sequence number $n$. Replicas accept the selected value $m$ for sequence number $n$ if the \textit{progress certificate} is valid and $m$ is selected as per \textit{progress certificate}. With this scheme, the protocol claims all correct replicas agree on a common value $v$ at a sequence number $n$. Below we show a simple scenario that breaks this claim.

%% file: sections/violation.tex
\section{Breaking Safety}
To demonstrate the issue, we adopt a similar explanation as presented in \cite{abraham2017revisiting}. Consider four replicas $i_1$, $i_2$, $i_3$, $i_4$ of which one, $i_1$, is Byzantine. All replicas participate in the agreement sub-protocol to decide on a common value for sequence number 1.

\begin{itemize}[\label{}]
\item \textbf{View 1: Primary $i_1$}
\begin{enumerate}[1.]
    \item In view 1, the primary $i_1$ sends a PREPARE message for value $a$ to replicas $i_2$ and $i_3$.
    \item Primary $i_1$ equivocates and sends conflicting PREPARE message for value $b$ to replica $i_4$.
    \item Replicas $i_2$ and $i_3$ accept the well-formed PREPARE message, send COMMIT messages to other replicas.
    \item Only Replica $i_3$ receives $2f+1$ identical COMMIT message for value $a$ (including Primary $i_1$'s PREPARE message for value $a$).
    
    Here, \textbf{for Replica $i_3$, value $a$ is committed at sequence number 1.}
    \item Replica $i_4$ also accepts well-formed PREPARE message for value $b$ and sends COMMIT messages for value $b$.
\end{enumerate}
At this stage, all further messages are delayed triggering a view-change.

\item \textbf{View 2: Primary $i_2$}
\begin{enumerate}[1.]
\item In view 2, primary $i_2$ collects VIEW-CHANGE messages from itself, $i_1$ and $i_4$ as follows:
 \begin{itemize}
     \item Replica $i_2$ sends its accepted value $a$.
     \item Replica $i_4$ sends its accepted value $b$.
     \item Replica $i_1$ (which is Byzantine) equivocates and sends value $b$.
 \end{itemize}
 Here, out of $2f+1$ VIEW-CHANGE messages, no \textit{commit certificate} exists and $f+1$ votes for value $b$. As per the specification, primary $i_2$ chooses $b$.
 \item Primary $i_2$ sends PREPARE messages to all replicas for value $b$.
 \item Replicas $i_1$, $i_2$ and $i_3$ accept well-formed PREPARE messages for value $b$ and send COMMIT messages for value $b$ to all replicas.
\item Replica $i_1$, $i_2$ and $i_4$ receive $2f+1$ identical COMMIT messages for value $b$. \textbf{For these replicas, value $b$ is committed at sequence number 1}.
\end{enumerate}
\end{itemize}